# Beyond the consensus: dissecting within-host viral population diversity of foot-and-mouth disease virus using next-generation genome sequencing

Running title: Intra-sample sequence diversity of FMDV

Caroline F. Wright[1,2,†], Marco J. Morelli[2,†,*], Gaël Thébaud[3], Nick J. Knowles[1], Pawel Herzyk[4], David J. Paton[1], Daniel T. Haydon[2], Donald P. King[1]

† These authors equally contributed to this work

1. Institute for Animal Health, Ash Road, Pirbright, Woking, Surrey GU24 0NF, United Kingdom

2. MRC – University of Glasgow Centre for Virus Research, Insitute of Biodiversity, Animal Health and Comparative Medicine, College of Medical, Veterinary and Life Sciences, University of Glasgow G12 8QQ, United Kingdom

3. Institut National de la Recherche Agronomique (INRA), UMR BGPI, Cirad TA A-54/K, Campus de Baillarguet, 34938 Montpellier cedex 5, France

4. The Sir Henry Wellcome Functional Genomics Facility, Faculty of Biomedical and Life Sciences, University of Glasgow G12 8QQ, United Kingdom

* Corresponding Author: Marco J. Morelli, Mailing Address: Graham Kerr Building, University of Glasgow G12 8QQ, United Kingdom, Email: Marco.Morelli@glasgow.ac.uk, Tel: 0044 141 330 6638, Fax: 0044 141 330 5971

Word count: Abstract 250. Main text 5767. Appendix 596.

## Abstract

The sequence diversity of viral populations within individual hosts is the starting material for selection and subsequent evolution of RNA viruses such as foot-and-mouth disease virus (FMDV). Using next-generation sequencing (NGS) performed on a Genome Analyzer platform (Illumina), this study compared the viral populations within two bovine epithelial samples (foot lesions) from a single animal with the Inoculum used to initiate experimental infection. Genomic sequences were determined in duplicate sequencing runs, and the consensus sequence determined by NGS, for the Inoculum, was identical to that previously determined using the Sanger method. However, NGS reveals the fine polymorphic sub-structure of the viral population, from nucleotide variants present at just below 50% frequency to those present at fractions of 1%. Some of the higher frequency polymorphisms identified encoded changes within codons associated with heparan sulphate binding and were present in both feet lesions revealing intermediate stages in the evolution of a tissue-culture adapted virus replicating within a mammalian host. We identified 2,622, 1,434 and 1,703 polymorphisms in the Inoculum, and in the two foot lesions respectively: most of the substitutions occurred only in a small fraction of the population and represent the progeny from recent cellular replication prior to onset of any selective pressures. We estimated an upper limit for the genome-wide mutation rate of the virus within a cell to be $7.8 \times 10^{-4}$ per nt. The greater depth of detection, achieved by NGS, demonstrates that this method is a powerful and valuable tool for the dissection of FMDV populations within-hosts.

# Introduction

RNA viruses evolve rapidly due to their large population size, high replication rate and poor proof-reading ability of their RNA-dependent RNA polymerase. These viruses exist as heterogeneous and complex populations comprising similar but non-identical genomes, but the evolutionary importance of this phenomenon remains unclear (15, 16, 25). Consensus sequencing identifies the predominant or major viral sequence present in a sample, but is uninformative about minority variants that are present. Evidence for population heterogeneity, where individual sequences differ from the consensus sequence, has been routinely obtained using cloning approaches (1, 8), providing insights into the evolutionary processes that shape viral populations. Unfortunately, these cloning processes are laborious and usually provide only a limited resolution of the mutant spectrum within a sample.

Next-Generation Sequencing (NGS) techniques offer an unprecedented 'step-change' increase in the amount of sequence data that can be generated from a sample. Albeit mostly used for de-novo sequencing of large genomes, NGS can be applied to re-sequence short viral genomes to obtain an ultra-deep coverage. Therefore, NGS has the potential to provide information beyond the consensus for a viral sample by revealing nucleotide substitutions present in only a small fraction of the population. Several studies have previously used the 454 pyrosequencing platform (Roche Applied Science) to detect minority sequence variants for human viruses such as HIV-1 (17, 22, 29, 36, 42, 45, 47), hepatitis B (32, 43), hepatitis C (48) and attenuated virus (46). A promising alternative to 454, is reversible terminator-based sequencing chemistry utilized by the Illumina sequencing platform (Genome Analyzer II). The lower costs of the runs and the higher throughput of this NGS approach are likely to make it widely used for deep-sequencing genomic investigations in the future (41). Illumina sequencing was recently used to obtain sequences of West Nile Virus, through virus-derived siRNA (3), mutant viruses of severe acute respiratory syndrome (14), and human rhinovirus (7).

The aim of this study was to explore the extent to which the Illumina sequencing platform can be used to characterize and monitor changes in viral sequence diversity that occurs during replication of a positive-stranded RNA virus within a host. This study uses NGS to dissect foot-and-mouth disease virus (FMDV) within-host population structure, at a depth unobtainable by previous cloning techniques. Belonging to the *Picornaviridae* family, FMDV is highly infectious causing vesicular lesions in the mouth and on the feet of cloven-hoofed animals. The samples analysed here were collected during an infection experiment, in which a bovine host was inoculated with FMDV. We developed a protocol that enabled identification of artifacts introduced during amplification and sequencing which was used to validate and quantify the minority sequence variants that were detected. In particular, we expected to see evidence for the reversion of capsid amino acid residues responsible for heparan sulphate (HS) binding associated with replication of a cell culture adapted strain of FMDV in a mammalian host (20, 37). Although this study was conducted using FMDV, we anticipate that the features we observe may be broadly representative of populations found in samples obtained from other positive-stranded RNA viruses.

# Materials and Methods

## *Sample preparation and genome amplification*

The samples analysed were collected during an infection experiment, in which a single bovine host was inoculated intradermolingually with a dose of $10^{5.7}$ 50% tissue culture infective doses ($TCID_{50}$) of FMDV ($O_1$BFS 1860). The full-length FMDV genome sequence of this sample had been previously determined using Sanger sequencing (EU448369) and was used as a reference genome in this study. The Inoculum was derived from a bovine tongue vesicle specimen that had been passaged extensively in cell culture (9).

Total RNA (TRIzol, Invitrogen, Paisley, UK) was extracted from a sample of the Inoculum as well as two 10% tissue suspensions prepared from epithelial lesions (front left foot [FLF] and back right foot [BRF]) collected from the animal at 2 days post inoculation. Reverse transcription was performed using an enzyme with high specificity (Superscript III reverse transcriptase, Invitrogen), and an oligo-dT primer (see Table 1). For each sample, two PCR reactions generating long overlapping fragments (4557 bp and 4317 bp respectively) were carried out using a proof-reading enzyme mixture (Platinum Taq Hi-Fidelity, Invitrogen). For biosecurity reasons these individual fragments together comprised <80% of the complete FMDV genome, and corresponded to nts 351-4909 and 3859-8176 of EU448369. This enabled the amplified DNA to be transported outside of the high containment FMD laboratory for sequencing at The Sir Henry Wellcome Functional Genomics Facility (University of Glasgow). The samples were amplified using the following cycling programme: 94 °C for 5 min, followed by 94 °C for 30 s, 55 °C for 30 s and 70 °C for 4 min, with a final step of 72 °C for 7 min. For each RNA sample, the number of PCR cycles used was optimized (using parallel reactions undertaken using Picogreen) such that products were collected from the exponential part of the amplification curve prior to the plateau phase. Once established for each sample, the same optimized cycle number was used for both runs. Individual PCR products were visualized using agarose-gel electrophoresis and quantified (Nanodrop, Labtech), after which the concentrations of each PCR fragment were adjusted to equimolar ratios for each of the three samples prior to sequence analysis. We repeated the PCR of the original reverse-transcribed sample in order to obtain an independent replica of the amplified sample. The number of viral RNA copies put into the initial PCR reaction was established by quantitative PCR for each of the samples (4).

### *Next Generation sequencing*

Sequencing was carried out on the Genome Analyzer II platform (Illumina). Briefly, DNA was fragmented using sonication and the resultant fragment distribution assessed by an Agilent BioAnalyzer 2100. After size selection of fragments between 300 and 400 bp, a library of purified genomic DNA was prepared by ligating adapters onto the fragment ends to generate flow-cell suitable templates. A unique 6-nt sequence index, or 'tag' for identification during analysis, was added to each sample by PCR. Once the adapter/index modified fragments were pooled and attached to the flow-cell by complimentary surface-bound primers, isothermal 'bridging' amplification formed multiple DNA 'clusters' for reversible-terminator sequencing, yielding reads of 50 nucleotides. We conducted two sequencing runs: in the first, we sequenced on a single lane the three amplified viral populations (Inoculum, FLF and BRF) after tagging. The second run was performed on a different flow cell: again, we sequenced the same populations on a single lane, using a second, independent amplification of the three original cDNAs. The second run was performed after the Illumina Genome Analyzer went through an upgrade and was able to deliver longer reads of 70 nucleotides.

*Data filtering*

In order to make direct comparisons between the two runs, we trimmed reads from the second run to 50nt. Typically, quality scores decreased along a read, as the reliability of the sequencing process decreased with the number of cycles of the Sequencing Platform. The second run yielded much better qualities following an upgrade of the Illumina platform. For both runs, reads with average error per nt below a fixed threshold ($\theta = 0.2\%$) were discarded to generate a flatter error profile along the read (see Appendix and Figure A1). The first and last 5 nts of each aligned read were removed from the analysis as they showed a higher number of mismatches to the reference sequence due to insertions or deletions close to the edges of the reads (Figure A2). More details can be found in Appendix.

*Validation and analysis of sequence diversity in the samples*

The frequency of site-specific polymorphisms was estimated from the frequency of mismatches of the aligned reads to the reference genome. A proportion of these mismatches was expected to be artifactual, arising from a base mis-calling in the sequencing process, or from a PCR error in the amplification of the sample. In order to identify polymorphisms arising from possible base mis-calls in the sequencing reaction, we used the quality score of each nucleotide read to compute the average probability of a sequencing error, $p_i$, at each site i. Typical values of $p_i$ are around 0.1%. Assuming sequencing errors to be independent, we computed the expected number of such errors as the mean of the binomial distribution $B(x; p_i, n_i)$, where $n_i$ is the coverage of site i. If the observed number of mismatches exceeded this expected number of errors in both runs then we excluded the possibility of a sequencing error. On the other hand, we hypothesize that the probability that PCR errors in both runs independently generated identical base changes at the same site is very low. Based on values quoted for the enzymes used, we estimated that the error rate for the combined RT-PCR amplification process to be $7.7 \times 10^{-6}$ per base pair copied (2, 31, 38). We therefore defined polymorphic sites that could not be attributed to sequencing errors and at which both the most common and second most common nucleotides were the same between the two runs to be 'qualitatively validated sites'. For each site in the set of qualitatively validated polymorphisms, we computed the 95% confidence intervals for the polymorphism frequency using the binomial distribution above. If the 95% confidence intervals from each run overlapped, we defined the polymorphism frequency estimates from the two runs to be in quantitative agreement.

We assessed the quantitative repeatability of site-specific polymorphism frequency estimates by calculating Spearman Rank correlation coefficients between polymorphism frequencies in the samples within each run and between polymorphism frequencies from runs 1 and 2.

We counted the number of transitions (Ts) and transversions (Tv) observed at qualitatively validated sites across the genome, we computed $\kappa=2Ts/Tv$, and the relative distribution of mutations across the 1st, 2nd, and 3rd codon positions across the open reading frame (ORF). We obtained an estimate for dN/dS as follows: for each codon of the reference ORF, we computed the expected number of synonymous ($s_i$) and non-synonymous sites ($n_i$) and, for each read j spanning that codon, the number of observed synonymous ($s^r_{ij}$) and non-synonymous substitutions ($n^r_{ij}$). Using all the codons where $s_i>0$, we then obtained an estimate of the number of synonymous substitutions per synonymous site ($p_S$) and non-synonymous substitutions per non-synonymous site ($p_N$): $p_S = \frac{1}{n_{cod}} \sum_{i=1}^{n_{cod}} \frac{1}{m_i} \sum_{j=1}^{n_i} \frac{s_{ij}^r}{s_i}$ (and analogously for $p_N$) ,where $m_i$ is the number of reads covering codon i and $n_{cod}$ is the total number of codons in the ORF. From $p_N$ and $p_S$ we obtained dN/dS according to (34).

We calculated the number of validated sites at which STOP-codons are observed within the reading frame, and used these count to estimate an upper limit on the mutation rate. Let $n_i$ be the coverage at the ith nucleotide position, and let $x_{i,obs}$ be the number of reads indicating a STOP codon at the ith position. Assuming independence, the probability density function describing the number of mutations, $x_i$, that might be observed at site i is the binomial $B(x_i;\lambda,n_i)$ where λ is the mutation frequency, corresponding to mutations accumulated at a site during a single cellular passage. The maximum likelihood estimate of λ is $\sum_i x_{i,obs} / \sum_i n_i$ (18). Using a flat conjugate prior distribution (beta function with shape parameters set to 1), we obtained confidence intervals for λ from the corresponding posterior distribution (beta function with parameters $1+\sum_i x_{i,obs}$ and $1+\sum_i (n_i - x_{i,obs})$ (21)). Assuming an equal probability for each mutation, λ is related to the mutation rate μ (per nucleotide, per single copying event) via the relation $\lambda = 2ga\mu$ (44), where g is the number of transcription generations (positive -> negative -> positive) the virus underwent in the cell. Here, we assume g=1, which corresponds to a stamping machine replication strategy and therefore to the minimum number of copying events in a cell. *a* is a factor weighting the fraction of mutations generating a STOP codon among all the possible changes that could arise at a single nucleotide position: we only consider sites whose mutations can lead to a STOP codon. Among the 18 codons that are one mutation away from a STOP, 5 of them (UCA, UUA, UAC, UAU, UGG) can reach a STOP codon through either two different mutations to the same position, or a single mutation to one of two different positions. Assuming the same probability for each of the 3 nucleotide mutations, we obtain then a = (4*2+15*1)/(3*19) = 0.4035.

Randomizations were conducted whereby we assembled putative ‘clones’ from the read data by sampling nucleotides randomly from (qualitatively validated) nucleotide frequencies observed at each site along the genome. We computed the median number of observed nucleotide substitutions (those differing from the consensus of the resampled clones) in sets of 26 independently such assembled clones and these numbers were compared with equivalent numbers from real clones obtained from an individual cow naturally infected with FMDV (8).

The complexity of the viral populations was characterized by computing the entropy of the viral populations: $S = -\frac{1}{N}\sum_{1=1}^{N} \sum_{j\in\{A,C,G,T\}} p_{ij} \ln p_{ij}$ where *N* is the number of sites and $p_{iX}$ is the fraction of reads bearing nt *X* at site *i*. The entropy measures the amount of “disorder” in the population, and it is maximum at a site when all four bases are equally represented.

## Results

In this section, we discuss the results of the Illumina sequencing of three FMDV populations: the Inoculum (field sample O1/BFS1860/UK/67, used to artificially infect a bovine host), and two lesions developed on two different feet of the host, 2 days after inoculation.

### *Description and filtering of Illumina data*

Sequences from the Illumina Genome Analyzer platform consist of a collection of several million short reads. Sequencing was repeated following independent amplification of cDNA generated through PCR. In the first run ~8% of the reads were discarded because of unresolved nucleotides or corrupted tags. In the second run, ~3% of the reads were discarded. Each nucleotide (nt) of each read is characterized by a quality score, which quantifies the reliability of the base-calling process during the sequencing. Only reads whose average error per nt was below 0.2% (66% for the first run and 95% for the second run) were considered for this analysis. Further details about the reads and the filtering process can be found in the Appendix.

### *Coverage and consensus genomes*

Reads that passed the quality test were aligned to the consensus genome sequence of the starting material from which the Inoculum was prepared (see Appendix). The mean coverage of the reference genome in the first run was 4863x for the Inoculum, 8665x for the Front Left Foot (FLF) sample and 6594x for the Back Right Foot (BRF) respectively, while for the second run it was 16827x for the Inoculum, 11924x for FLF and 15945x for BRF (Figure 1A, B). For some samples (Inoculum and BRF, first run and FLF, second run), the coverage for the two PCR fragments composing the viral genome was not equal. More details on the statistics of the Illumina yield can be found in the Appendix.

We obtained consensus genomes for each sample, by identifying, site by site, the most abundant nucleotide in the aligned reads. As expected, the consensus for the Inoculum exactly matched the reference genome at all sites. For FLF, both runs indicated two substitutions (nt 2767, G–>A, and nt 8140, G–>T). For BRF sample, the two runs suggested slightly different consensus sequences: the first run revealed five substitutions (nt 2767, G–>A, nt 3138, G–>A, nt 5138, T–>C, nt 7354, C–>T, nt 8134, C–>T), whereas the second run had none. However, at position 8134 about 30% of the reads in the second run showed a T in place of a C, and at position 2767 5% of the reads had an A in place of a T. At the remaining 3 sites, the second run had a small number of reads confirming the polymorphism found in the first run. This result indicates that the same pattern of variation is present in both runs, although the frequency of the mutations is not in quantitative agreement across the two runs for BRF. Finally, the second run showed an almost-consensus substitution in 49.9% of the reads (nt 2754 C->T), which was present at a 10% frequency in the first run.

### *Validation of polymorphic sites*

Mismatch frequencies, obtained by showing site by site the fraction of reads differing from the consensus genome, are shown in Figure 2 (first run) and Figure 3 (second run). An evident correlation is present between the regions of the sample genomes receiving low coverage and those with the largest fraction of sites showing no variation (Figs. 2A, second half, 2C, first half and 3B, first half). Using these raw data, and considering only sites receiving coverage of 100x or more, we found polymorphisms at 7,755 sites in the Inoculum, 7,730 in FLF and 7,710 in BRF, out of the 7,825 nt sequenced. While a few sites exhibited higher levels of polymorphism, the vast majority of sites displayed a mismatch frequency around 0.1%.

After screening for possible PCR and sequencing artefacts, we found that qualitatively validated polymorphisms were present at 2,622 sites for the Inoculum, 1,434 in FLF and 1,703 for BRF. The different consensus genomes obtained for BRF in the two runs can be in part reconciled by noting that all six substitutions observed (nt 2754, 2767, 3138, 5138, 7354, 8134) are qualitatively validated in each run. We observed 2,469 quantitatively validated sites in the Inoculum (94% of qualitatively validated sites), 1,303 sites from the FLF (91% of qualitatively validated sites) and 1,528 sites (90% of qualitatively validated sites) from the BRF

Site-specific polymorphism (SSP) frequency at qualitatively validated sites was correlated between the two runs for each of the three samples (Figure 4). The intra-run correlation for run 1 (Spearman Rank correlation: 0.64 [Inoc-FLF], 0.55 [Inoc-BRF] and 0.60 [FLF-BRF]) was higher than run 2 (Spearman Rank correlation: 0.40 [Inoc-FLF], 0.43 [Inoc-BRF] and 0.42 [FLF-BRF]). The reason for the poor intra-run correlation for run 2 is unclear. The number of viral RNA copies present in the initial PCR reactions was found to be large ($3.2 \times 10^9$ for the Inoculum, $6.4 \times 10^8$ for FLF and $2.4 \times 10^8$ for BRF): assuming that the PCR process amplifies all genomes with the same probability, the probability of resequencing the same genome is exceedingly low ($<10^{-5}$), thus excluding the possibility of biases due to low viral load in the RNA. However, relative to run 1, run 2 yielded consistently lower amounts of DNA library concentrations per sample prior to sequencing (3.4 vs 4.9 ng/µl, 3.7 vs 10.6 and 3.4 vs 9.5 ng/µl ng/µl for the inoculum, FLF and BRF respectively): and this may have contributed to the differences between the two consensus sequences. Despite the discrepancy in frequency of these mutations between runs, the fact that the same mutations are present at the same qualitatively validated sites in both runs provides confidence that the mutations are genuine and not artifacts. The intra-run correlation, together with the high fraction of quantitative validation among the qualitatively validated SSPs provides sound evidence that nt changes are linked between the different samples. Inter-run correlation between the samples (Spearman Rank correlation: 0.34 vs 0.44 and 0.50) indicates that validated polymorphisms are unlikely to be artifacts.

### *Distribution of polymorphisms across the genome*

There were 12 SSPs, whose average frequency between the two runs is above 1% in the Inoculum, 19 in FLF, and 25 in BRF (see Supplementary Table). Some of these were clustered in the capsid protein region (beginning of protein VP3) (1 in the Inoculum, 4 in FLF and 5 in BRF) and in the 3' untranslated region (UTR) (6 in the Inoculum, 5 in FLF and 6 in BRF). Where single reads spanning these sites within the VP3 or 3'UTR were available, there was no evidence that that these mutations were linked together on individual FMDV genomes. In particular, the first cluster was shared between the two foot samples and corresponded to changes encoding amino acid residues associated with heparan sulphate (HS) binding. The Inoculum used in this experiment had undergone extensive cell culture passage and, in common with other in-vitro adapted viruses, utilizes HS as a cellular receptor (27). Subsequent replication in mammalian hosts drives the reversion of positively charged amino acid residues at specific sites in the viral capsid (20, 37). A consensus level substitution (>50%) exists within both feet samples of run 1 compared to the reference sequence (see above and the Supplementary Table). This polymorphism corresponded to a change within the 60$^{th}$ codon of protein VP3 ($VP3^{60}$). Although below the level of the consensus sequence, additional qualitatively validated SSPs that were present in both feet samples were detected at four further sites ($VP2^{134}$, two codon positions within $VP3^{56}$ and $VP3^{59}$) that impact on the ability of FMDV to bind HS. All but one of the mutations clustered within the 3' UTR of the three samples were located within the first four RNA-RNA pairings either side of the apex of a conserved stem-loop. This structure, one of two stem-loops previously predicted for FMDV and other picornaviruses (6) (33) is thought to generate long-distance RNA-RNA interactions that may impact upon viral replication (40). The presence of shared mutations between the two foot samples suggests a common history for the viruses arising as a result of the shared route of intra-host transmission from initial replication sites in the tongue to epithelial sites in the feet via the blood. However an alternative explanation – that the virus is subject to a common selection pressure in both sites cannot be ruled out.

## *Frequency of site-specific polymorphisms*

Some variability was present almost everywhere on the genome. Above minimum coverage of 100x, only 61 sites exhibited no polymorphism (0.79%) in the Inoculum, 59 (0.76%) in FLF and 49 (0.64%) in BRF. These sites received relatively low coverage, suggesting that the absence of observed genetic variability may be due to lack of power to detect it. By grouping the site-specific polymorphism frequencies into discrete bins, we can examine the proportion of sites experiencing different polymorphic frequencies and thereby obtain a comprehensive picture of the heterogeneity in the viral populations (Figure 5). Across the three samples, most sites exhibit a range of low-frequency SSPs between 0.01% - 1%. Only a few sites showed higher frequency polymorphism, and these sites were more numerous for the samples from the feet than from the Inoculum, indicating the generation of new high-frequency substitutions during the host passage. The dashed lines (Figure 5) correspond to the same analysis restricted to qualitatively validated sites and reveal a similar pattern.

## *Statistics of polymorphic sites*

NGS provided sufficient resolution to detect polymorphisms where two alternative substitutions are simultaneously present. The secondary substitutions (the third most abundant nucleotides in the reads at any particular site) that would have been qualitatively valid even in the absence of the second most abundant nucleotide substitution were present in 67 sites in the Inoculum, 15 in FLF and 41 in BRF. Secondary substitutions typically appear at frequencies below 0.5%, confirming the large amount of low-frequency variability in the samples.

Table 2 shows that transversions are rare among the validated mutations, and thus $\kappa$ (defined as 2Ts/Tv) is high (however, similar values were reported in (8)). The ratio of non-synonymous to synonymous substitutions in the open reading frame, dN/dS, is higher for FLF than for the other two samples because of the presence of the non-synonymous mutations in a large number of reads at positions 2754 and 2767, associated with heparan sulphate binding amino acid reversions within $VP3^{56}$ and $VP3^{60}$ respectively. The mutation frequency at the third codon position is only marginally higher than in the first and second positions. Taken together, these observations suggest that the observed polymorphisms are dominated by mutations arising during the last round of intra-cellular replication and that have not been subject to extensive purifying selection. Further evidence of this lack of selective pressure is provided by the presence of validated polymorphisms generating STOP codons within the ORF. These mutations are clearly lethal for the virus and would therefore be removed from the population during infection of another cell: they must therefore have arisen during the most recent rounds of viral replication. STOP codons were found at 24 sites in the Inoculum, 9 sites in FLF and 21 sites in BRF, mostly at frequencies around 0.1% (with a single exception in BRF where a mutation generating a STOP codon is present in 0.7% of the reads).

The presence of STOP codons can be used to obtain an upper limit on the mutation rate (per nucleotide per transcription event) of this virus. We hypothesise that these mutations are lethal and are therefore generated in the last round of cellular replication. Moreover, by assuming a replication strategy involving the minimum number of copying events in the cell (the "stamping machine" strategy, see (44)), we obtain an upper bound for the mutation rate of $\mu = 7.8 \times 10^{-4}$ per nucleotide per transcription event (95% CI: $7.4 \times 10^{-4} - 8.3 \times 10^{-4}$), in line with previous estimates (e.g. (12, 13, 39)).

Finally, we can ask whether these results are broadly consistent with those acquired from cloning studies. In ref. (8), Cottam et al. generated 26 viral capsid clones from an FMDV sample taken from a single lesion of a bovine host. We simulated 10,000 sets of 26 viral capsid 'clones', essentially bootstrapping from the nucleotide frequencies revealed by the NGS alignments to be present at each site within the capsid genes. Of these 26 clones, the median number of sequences in each of the 10,000 simulated data sets that were identical to the consensus was 12 (95% CI: 5-17), compared to 15 observed in ref. (8). The median number of simulated clones containing 1, 2, 3, and 4 differences compared to the consensus were 9 (95% CI: 4-14), 3 (95% CI 1-7), 1, (95% CI: 0-3), and 0 (95% CI: 0-1) respectively. These numbers correspond well with those obtained by Cottam et al., (8) which were 6, 3, 2, and 0 respectively.

### *Complexity of the viral populations*

In the host, the viral population evolves via extensive replication, mutation, and, at the same time, selection. The result of these combined processes can be quantified by computing how much 'diversity' is present within the three samples, using an entropy-like measure $S$ that, site by site, takes a maximum value when all nts are present in the same proportion. The entropy of the three populations, computed over the qualitatively validated sites, shows higher values for the feet than for the Inoculum ($S$=0.01138 for FLF, $S$=0.01198 for BRF and $S$=0.00841 for Inoculum), suggesting that repeated cycles of cellular replication during passage in the host does result in greater viral population diversity relative to the Inoculum.

## Discussion

This study describes a novel use of Illumina NGS to investigate the population genetic structure of a positive-stranded RNA virus causing an acute-acting disease in hosts. These experiments generated an unprecedented amount of sequence data and required a new systematic approach to confidently distinguish between sequences that were actually present in the samples from artifacts introduced during the amplification and sequencing steps of the sample processing. Results obtained here were consistent with the findings of previous investigations, providing validation on the use of NGS in the study of FMDV evolution within a host: Carrillo et al. (5) reports an average of 1-5 substitutions per animal passage during an infection experiment in pigs, in line with the 2 substitutions we found in FLF. However, the case of the BRF points out a more complex scenario that could not have been observed with consensus sequences only: the drift of mutations above and below the threshold needed to appear in the consensus. Apparent loss and subsequent regain of mutations during the transmission of the infection across hosts (5) can be explained with this mechanism, which is made more accessible to study by NGS. Moreover, the statistical characteristics of the SSPs we identified ($\kappa$, dN/dS) are very similar to those found previously (8), further corroborating the validity of our results. Finally, randomizations of the diversity measured in the capsid region allowed us to obtain simulated clones whose characteristics in terms of mutation were analogous to those found in (8). We conclude that NGS data can be used to examine the nucleotide diversity of each genome position at unprecedented resolution. Observing the mutant spectrum of the viral population at a fine resolution will provide a more sophisticated understanding of evolutionary processes shaping its variability.

Comparisons between the sequences recovered from the Inoculum and clinical lesions provide new insights into the impact of early replication events on viral evolution within a host. This study reveals that only a few sites displayed mutations present in a large fraction of the population, i.e. high frequency polymorphisms (>1%), while the vast majority of the polymorphisms were present at lower frequencies. We hypothesize that the high frequency polymorphisms have been selected over multiple rounds of replication within cells, and that the lower frequency polymorphisms most likely directly reflect the high rate of mutation experienced by these viruses, as our estimate of an upper limit of the genome-wide mutation rate suggests. In this study we used a cell culture adapted virus (as the Inoculum) which gave us the opportunity to monitor changes at specific loci associated with the HS binding site that were under selection pressure during initial replication in a mammalian host. Examination of these sites (collated in the Supplementary Table) reveals for the first time the presence of intermediate stages in the evolution of the viral population between a tissue culture adapted genome and a host-adapted genome.

Cordey et al. (7) investigate the dynamics of Human Rhinovirus (HRV) during an infection experiment and in HeLa cells, and find results similar to ours in terms of number of mutations fixed at the consensus level. However, while their approach identifies hot and cold spots in the HRV ORF, and some minority variants, the resolution is not sufficient to observe the micro-evolutionary processes whose signature lies in small fractions of the viral population (<2%). Moreover, their estimation of the substitution rate during the infection is based solely on the count of the nucleotides changed among those analyzed: although the value is compatible to our genome-wide, mutation rate, we believe that considering the cellular process of viral replication (and specifically assuming the minimum number or copying events in a cell) allows us to gain a better insight of the process generating variation in the viral population and obtain a more stringent upper bound.

Figure 5 reveals that the viral population sequences are highly heterogeneous supporting the findings of previous studies that have used cloning approaches (10, 28). However, the massively increased coverage enabled by NGS enables the nature of this heterogeneity to be established at much greater resolution. This is important for understanding viral evolutionary processes because heterogeneity is a necessary but not sufficient condition (23, 24) for the dominance of quasi-species dynamics [see (10, 15) and references therein]. For quasi-species dynamics to dominate the micro-evolutionary process, the frequency of a dominant sequence must be maintained primarily by the back-mutation or recombination of closely related genetic variants, rather than the faithful replication of any single genome. This requires a balance of two qualities: genetic variants closely related to the dominant sequence must be maintained at sufficiently high prevalence; and that mutation and recombination rates must be sufficiently high to generate the observed prevalence of the dominant sequence from these variants. Previous studies have examined this question empirically and concluded that these conditions are indeed met in many RNA viruses, mostly through studies of mutational robustness as a selectable trait ("survival of the flattest" effect in which selection acts not on the dominant sequence, but on the swarm of viruses immediately mutationally adjacent to the dominant sequence) (11) (35) and reviewed in (19) with particular focus on Hepatitis C virus. However, taking FMDV as an example, given that there are ~25,000 one-step mutant variants to any one sequence (3 alternative nucleotides at each position of the ~8,300 nucleotide genome), NGS approaches are clearly a powerful tool for examining directly whether viral populations are structured in a way that is consistent with a quasi-species dynamic.

NGS data can be coupled to evolutionary models to estimate parameters, such as the genome-wide mutation rate of FMDV. Here, we computed this number hypothesizing that the viral replication strategy followed the so-called "stamping machine" mode of replication, where all viral genomes leaving the cells are obtained as copies of "first generation" negative stranded genomes, which are in turn direct copies of the genomic RNA originally infecting the cells. For this reason, the estimate of $7.8 \times 10^{-4}$ per genome per duplication round should be considered an upper bound on the mutation rate which is a tighter estimate than previous figures obtained for other RNA viruses (12, 13) as a result of the deep coverage that NGS generates. Were the replication strategy "geometric" (i.e. including the possibility of several rounds of positive/negative strand copying before exiting the cell), the mutation rate would be several-fold (perhaps 3-6 times) lower (44). The assumptions that all nucleotide mutations at a site are equally likely, and that all STOP codons are generated by a *de novo* mutation are also likely lead to an overestimation of the mutation rate.

To present date, the analysis of the amount of complexity carried by a genome has mostly coincided with information-theoretical measures, aimed to quantify at the entropy and the frequency distributions of short oligomers (26, 30). This approach looks at the "horizontal" complexity along a genome; with NGS we are now able to obtain the closely-related sequences for a whole viral population in a single experiment, thus enabling us to look at the "vertical" complexity of the viral variants, i.e. at the amount of variability present in the population at each site.

A viral population within a host undergoes complex processes, including the onset of infection, cellular replication, selection, and migration to different tissues. In particular, it is not clear how the diversity generated within a cell propagates through a host to give rise to the amount of diversity we observe. The data collected in studies like this can be used for building models aimed at understanding the link between the micro-evolution of FMDV at the cellular scale with the population heterogeneity at the host scale. We anticipate that a model of viral replication across several cell generations within a host will produce a more stringent upper bound to the genome-wide mutation rate.

Although further work is required, these findings strongly suggest that data generated through the use of this methodology can provide novel insights into viral evolutionary dynamics at a greater resolution than previously achieved for a positive-stranded virus such as FMDV. In particular, the genome wide assessment of polymorphic frequencies is likely to be an important asset in the parameterization of models that can evaluate the role of quasi-species dynamics in RNA virus evolution.

## Acknowledgments

This work was supported by the Biotechnology and Biological Sciences Research Council, United Kingdom via a DTA PhD studentship, project BB/E018505/1, a SYSBIO postdoctoral grant, project BB/F005733/1, and the IAH's Institute Strategic Programme Grant on FMD. The authors would like to thank Dr N. Juleff for providing the samples used in the study and J. Galbraith at the Sir Henry Welcome Functional Genomics Facility, University of Glasgow for sequencing the samples and for providing advice. We would also like to thank C. Randall, L. Fitzpatrick and M. Jenkins for their care of the animals used in the experimental infection study and Derek Gatherer and Gareth Elvidge for useful discussions.

## Appendix: Statistics and treatment of Illumina sequences

### *Basic statistics of Illumina yield*

The reads obtained with the Illumina Genome Analyzer were collected in *.fastq* files. The first run consisted of a total of 7,190,884 reads of 57-nucleotides (nt) in length. The last 7 nts of each read defined the sequence tag, and were used to assign individual reads to each sample. Reads containing at least one unresolved nt (387,809, 5.55% of the total), and reads having a corrupted tag (207,749, 2.89% of the total) were removed from the analysis. The 6,595,326 remaining, 50nt-long reads, were assigned to the three samples: 1,723,151 (26.1%) had the first tag (corresponding to the Inoculum), 2,751,260 (41.7%) had the second tag (lesion on the Front Left Foot, or FLF) and 2,112,932 (32.0%) had the third tag (lesion on the Back Right Foot, or BRF).

The second run yielded 10,116,147 79-nt long reads, with the last 9 nts containing the sequence tag. 26,428 (0.27%) reads contained at least one unresolved nt and 288,230 (2.85%) reads had a corrupted tag and were removed from the analysis. Among the remaining 9,801,489 70-nt reads, 3,775,685 (38.5%) belonged to the Inoculum, 2,542,913 (25.9%) to FLF and 3,482,891 (35.5%) to BRF.

### *Data filtering*

The quality scores associated with each nucleotide were lower on the first run and decreased towards the end of reads (Figure A1). In order to make direct comparisons between the two runs, we trimmed reads from the second run to 50nt. Typically, quality scores decreased along a read, as the reliability of the sequencing process decreased with the number of cycles of the Sequencing Platform. As Figure A1 shows, a trade off is present between the number of reads kept and their quality. For both runs, we discarded reads with average error per nt below $\theta = 0.2\%$, resulting in a flatter error profile along the read.

With this choice of the threshold, 66% of the reads were retained from the first run (a total of 4,361,101 reads: 1,060,906 for the Inoculum, 1,736,381 for FLF and 1,328,588 for BRF), and 95% of reads from the second run (a total of 9,277,876 reads: 3,567,541 for the Inoculum, 2,412,897 for FLF and 3,303,438 for BRF). The better performance of the second dataset has to be attributed to an upgrade of the Illumina platform.

### *Reads alignment and trimming*

The vast majority of the filtered 50nt-reads aligned unambiguously with less than 5 mismatches to the reference inoculum genome, previously established using conventional Sanger sequencing (9) (Genbank accession no. EU448369), (run 1: 92.5% for Inoculum, 98.9% for FLF, and 97.8% for BRF, run 2: 95.8% for Inoculum, 98.4% for FLF and 96.2% for BRF). The remaining reads were either ambiguously aligned reads or contained a large number (>4) of mismatches to the reference sequence, and were discarded from the analysis. For each sample, an almost equal number of reads were derived from positive and negative strands of the viral cDNA.

Further filtering of the data was undertaken after alignment of the reads. Within the aligned reads, mismatches occurred with similar frequency at each of the 50nt of the reads, except from the edges, where a higher number of mismatches was observed (Figure A2). Presumably, these peaks were due to a small number of sequences with insertions or deletions close to the ends of the reads: for subsequent analysis we trimmed away the first and last 5 nts of each aligned read, leaving only the 40 central nucleotides where the mismatch curve was flat.

### *Data handling*

All data handling was performed with parsing scripts, written in C language, acting on plain text files.

TABLE 1: Primers. Oligonucleotide primers used for the amplification of the two large, overlapping FMDV genome fragments studied (omitting the S fragment up to and including the poly(c) tract), for both the first and second run

| PCR Set | Primer * | Primer Sequence (5' to 3') | Location on Genome + | Amplicon Size (bp) | Overlap (bp) |
|---|---|---|---|---|---|
| 1 | BFS-370F | CCCCCCCCCCCCCTAAG | 351-366 | 4557 | 1051 |
| | BFS-4926R | AAGTCCTTGCCGTCAGGGT | 4891-4909 | | |
| 2 | BFS-3876F | AAATTGTGGCACCGGTGA | 3859-3876 | 4317 | |
| | BFS-8193R | TTTTTTTTTTTTTTGATTAAGG | 8155-8176 | | |
| - | UKFMD/Rev6 | GGCGGCCGCTTTTTTTTTTTTTTTT | poly(A) | | |

* Last letter indicates a forward or reverse primer

+ Numbering according to Genbank sequence EU448369

TABLE 2: Statistics of polymorphic sites. General statistics of qualitatively and quantitatively validated SSPs receiving coverage larger than 100x. Ts: transitions in SSPs, Tv: transversions in SSPs, κ=2Ts/Tv, dN: non-synonymous mutations in the ORF, dS: synonymous mutations in the ORF, 1st, 2nd and 3rd: mutations in codon positions in the ORF.

| | Sites | SSPs | Ts | Tv | κ | dN/dS | 1st | 2nd | 3rd |
|---|---|---|---|---|---|---|---|---|---|
| Inoc | 7755 | 2622 | 2562 | 60 | 85.40 | 0.651 | 0.288 | 0.286 | 0.427 |
| FLF | 7730 | 1434 | 1400 | 34 | 82.36 | 1.065 | 0.326 | 0.333 | 0.341 |
| BRF | 7710 | 1703 | 1649 | 54 | 61.08 | 0.680 | 0.334 | 0.307 | 0.359 |

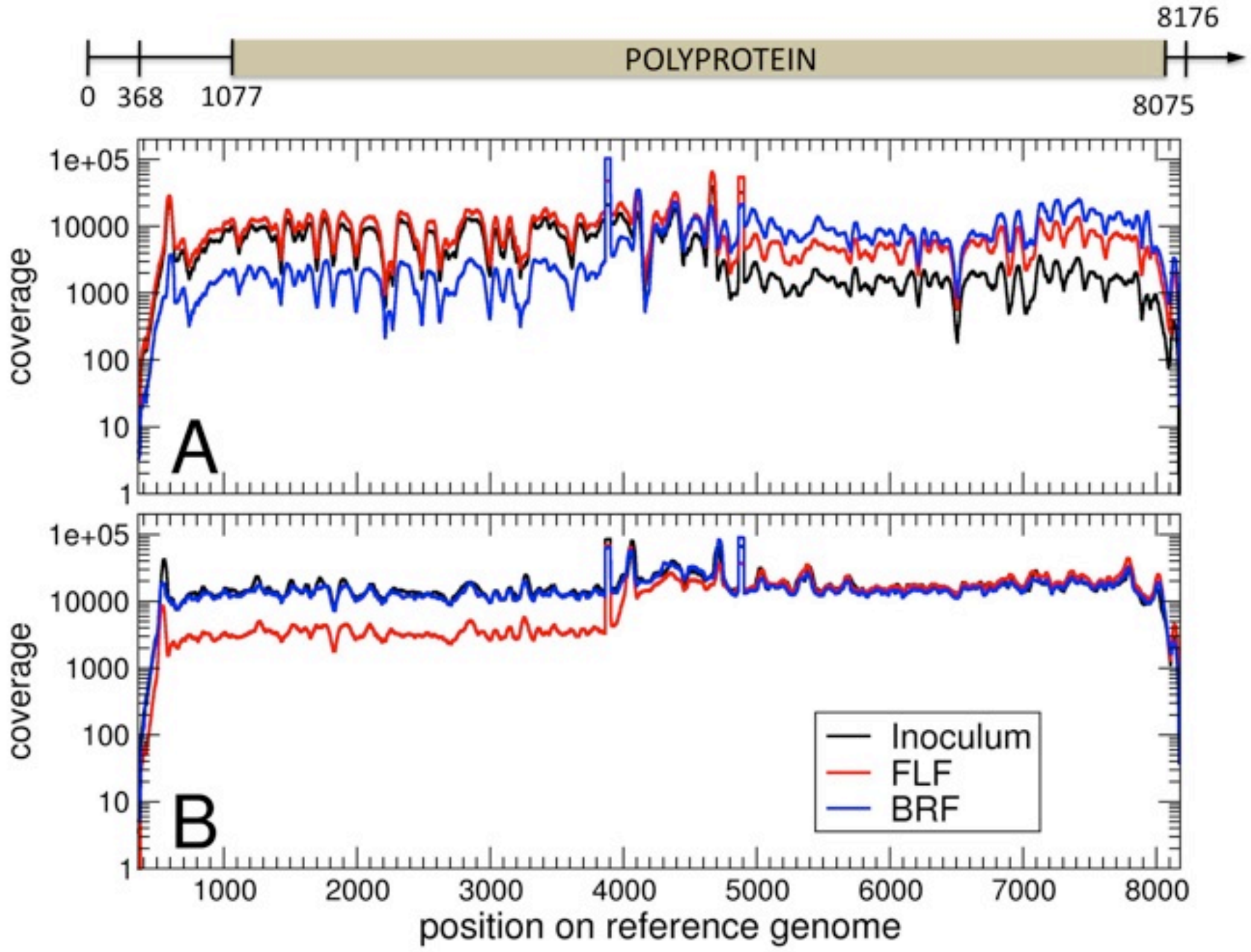

FIG. 1: Coverage of the reference genome. Obtained with the filtered, trimmed reads. Panel A: first dataset, panel B: second dataset. The three samples (Inoculum, Front Left Foot and Back Right Foot) receive a generous coverage from both runs, while fluctuations are higher on the first dataset. Average coverage is 4873x (Inoculum), 8665x (FLF) and 6594x (BRF) for the first dataset, and 16827x (Inoculum), 11924x (FLF) and 15945x (BRF) for the second dataset. On top of the figure, the sequenced fraction of the genome (nt 368-8176) is represented, together with the position of the polyprotein.

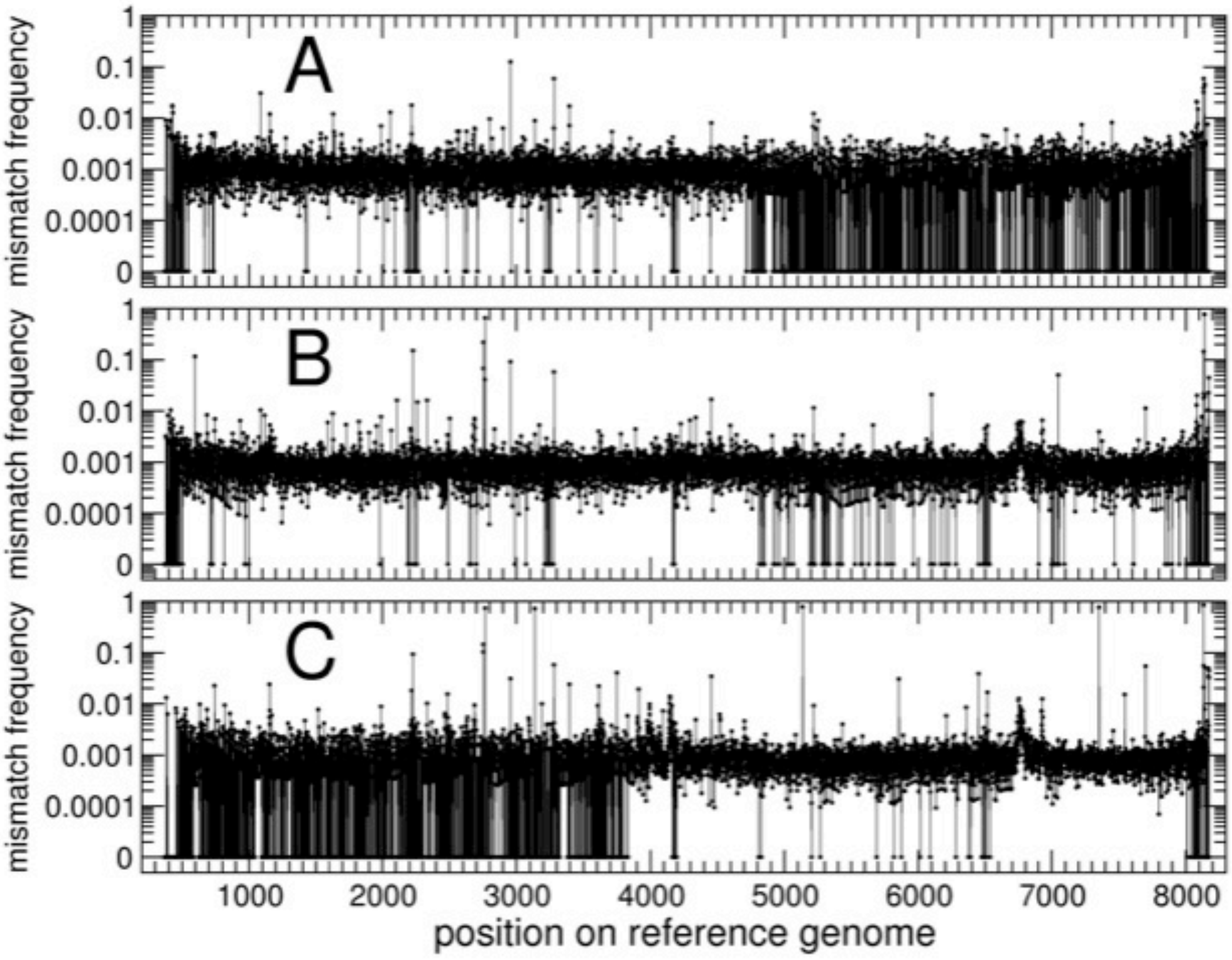

FIG. 2: Frequency of mismatches (first dataset). Obtained by aligning the reads to the reference genome. Panel A: Inoculum, panel B: FLF, panel C: BRF. The average mismatch frequency lies around 0.1% for all the three samples. At few sites, the mismatch frequency is higher; as expected, the number of these peaks is larger in the FLF and BRF than in the Inoculum. A small fraction of sites show perfect agreement of all the reads with the reference genome (mismatch frequency = 0).

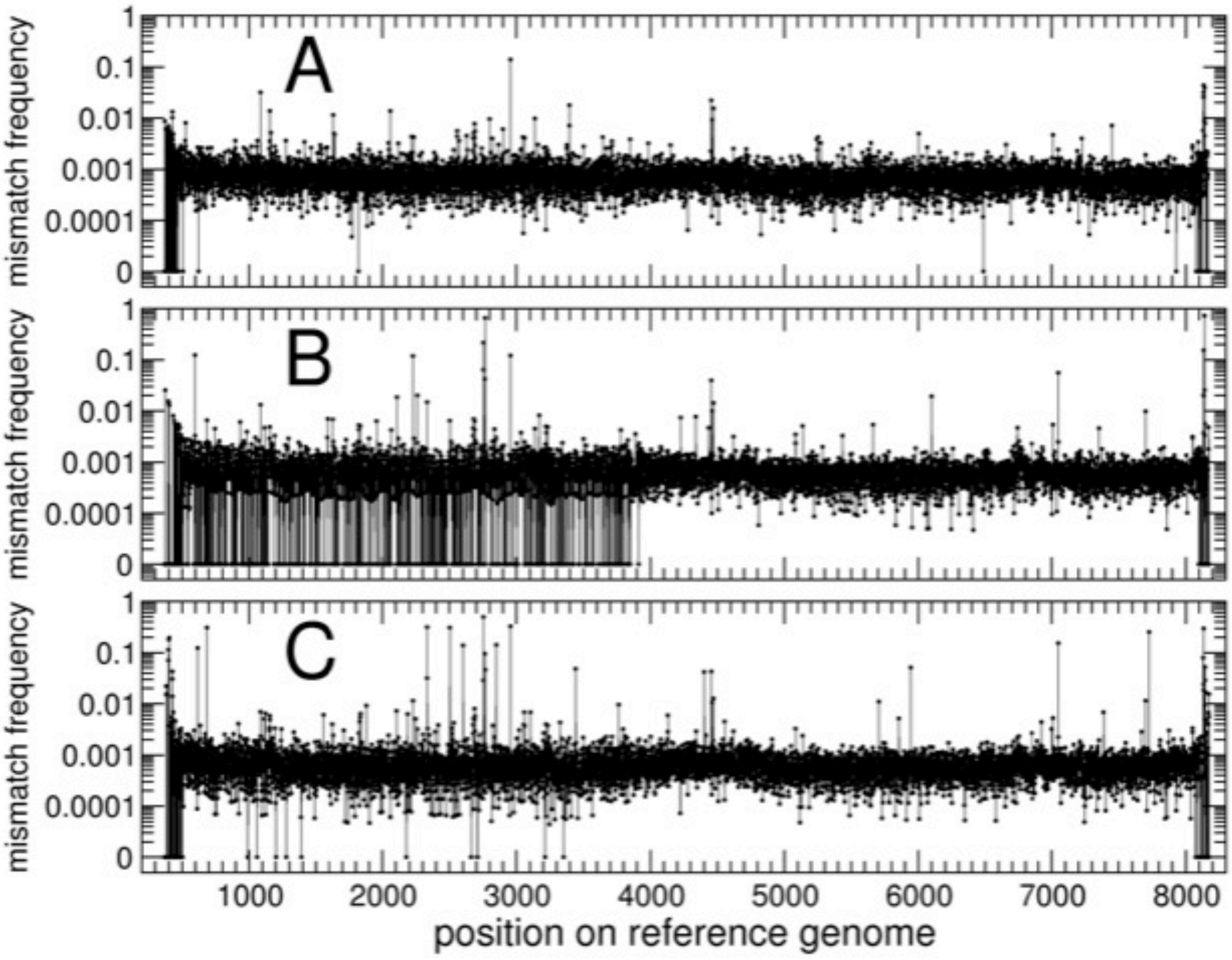

FIG. 3: Frequency of mismatches (second dataset). Obtained by aligning the reads to the reference genome. Panel A: Inoculum, panel B: FLF, panel C: BRF. This second dataset has higher coverage than the first one, and a lower fraction of sites with no mismatches. The average mismatch frequency is very similar to that of the first dataset.

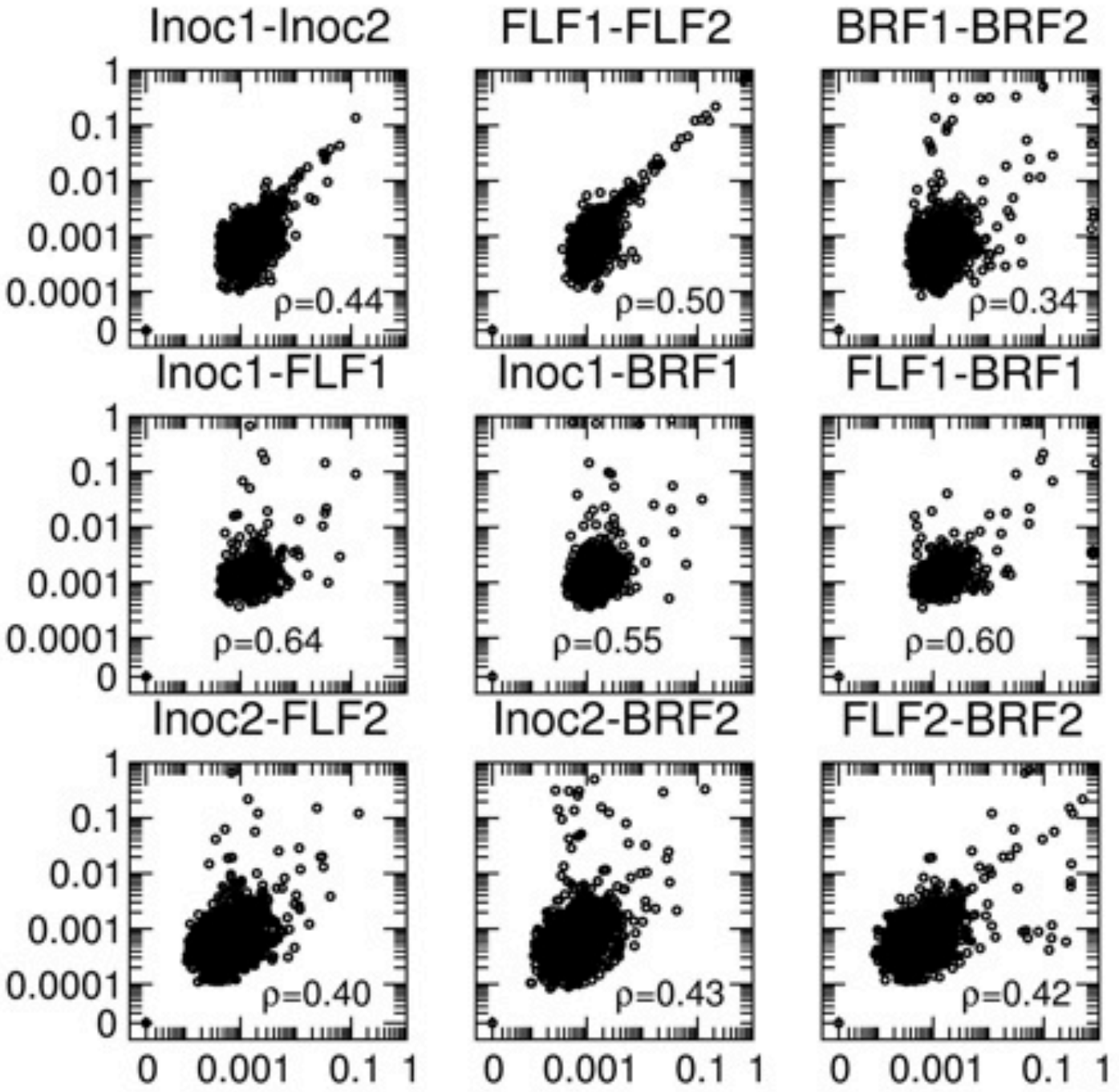

FIG. 4: Correlations of polymorphism frequencies in the viral populations. Correlations were computed between the two runs (first row) and within each run (second and third row). The Spearman rank correlation ρ is indicated for each pair of datasets. Only qualitatively validated SSPs receiving coverage above 100x in both runs are shown. The correlation coefficients between the two runs in the Inoculum and FLF are similar, while they are lower for BRF. The remaining panels show that the first run is more correlated than the second.

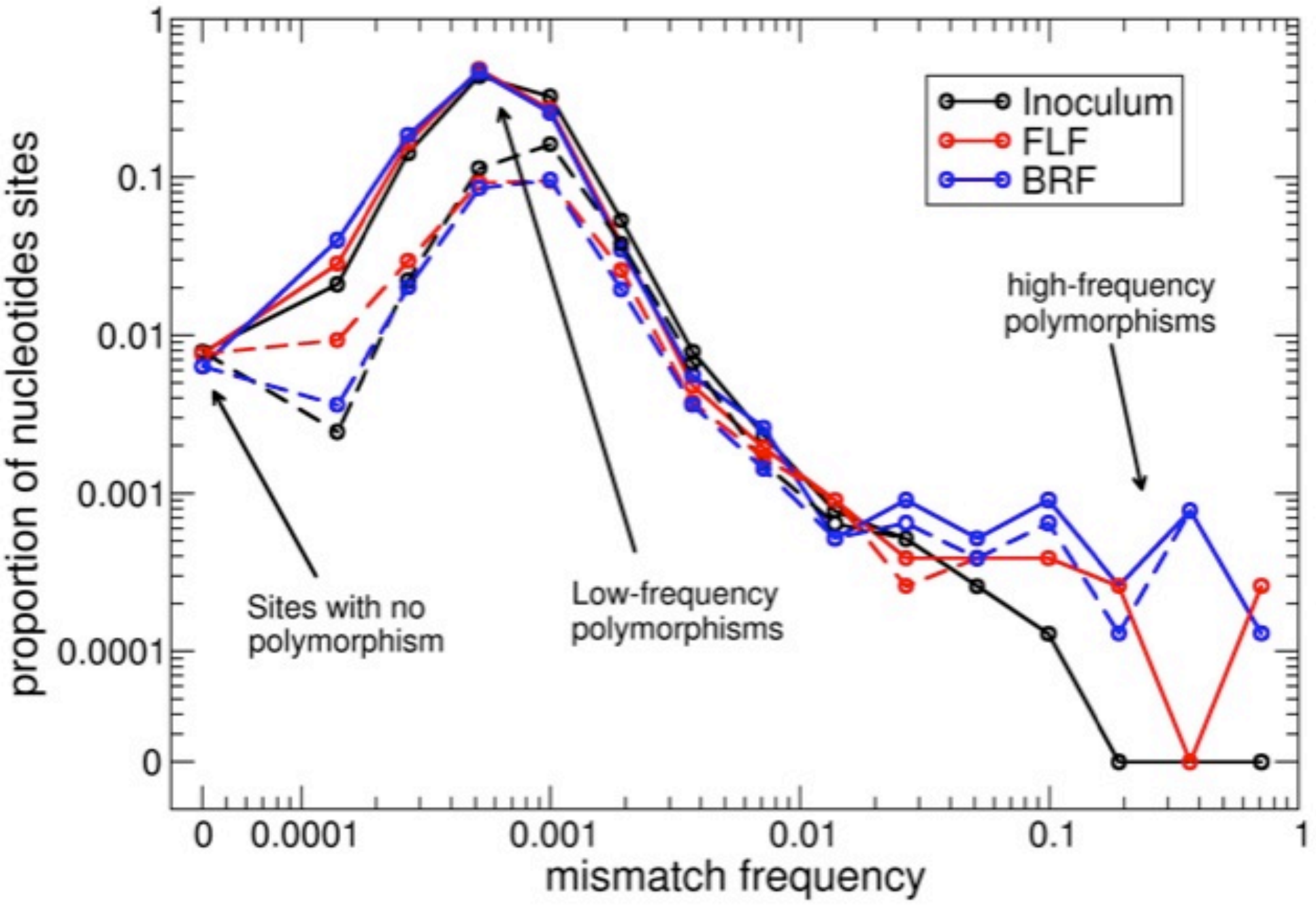

FIG. 5: Variability in the viral populations. Frequency distribution of the weighted averaged mismatch frequencies between the two runs, for the three samples (the ordinate represents the frequencies of sites showing that fraction of mismatches). Solid lines: all sites receiving minimum coverage of 100 in both runs (7,755 sites for Inoculum, 7,730 for FLF and 7,710 sites for BRF). Dashed lines: sites receiving coverage of 100 or more in both runs, and classified as validated site specific polymorphisms (SSPs) (2,622 sites for Inoculum, 1,434 for FLF and 1,703 for BRF). All lines show a similar trend: a small fraction of the sites (<1%) display no variability in both runs, most of the sites show a very mild amount polymorphism in the viral population (between 0.01% and 1%), while a very small fraction of the sites (0.14% for Inoculum, 0.22% for FLF and 0.39% for BRF) present variation at a level above 1%.

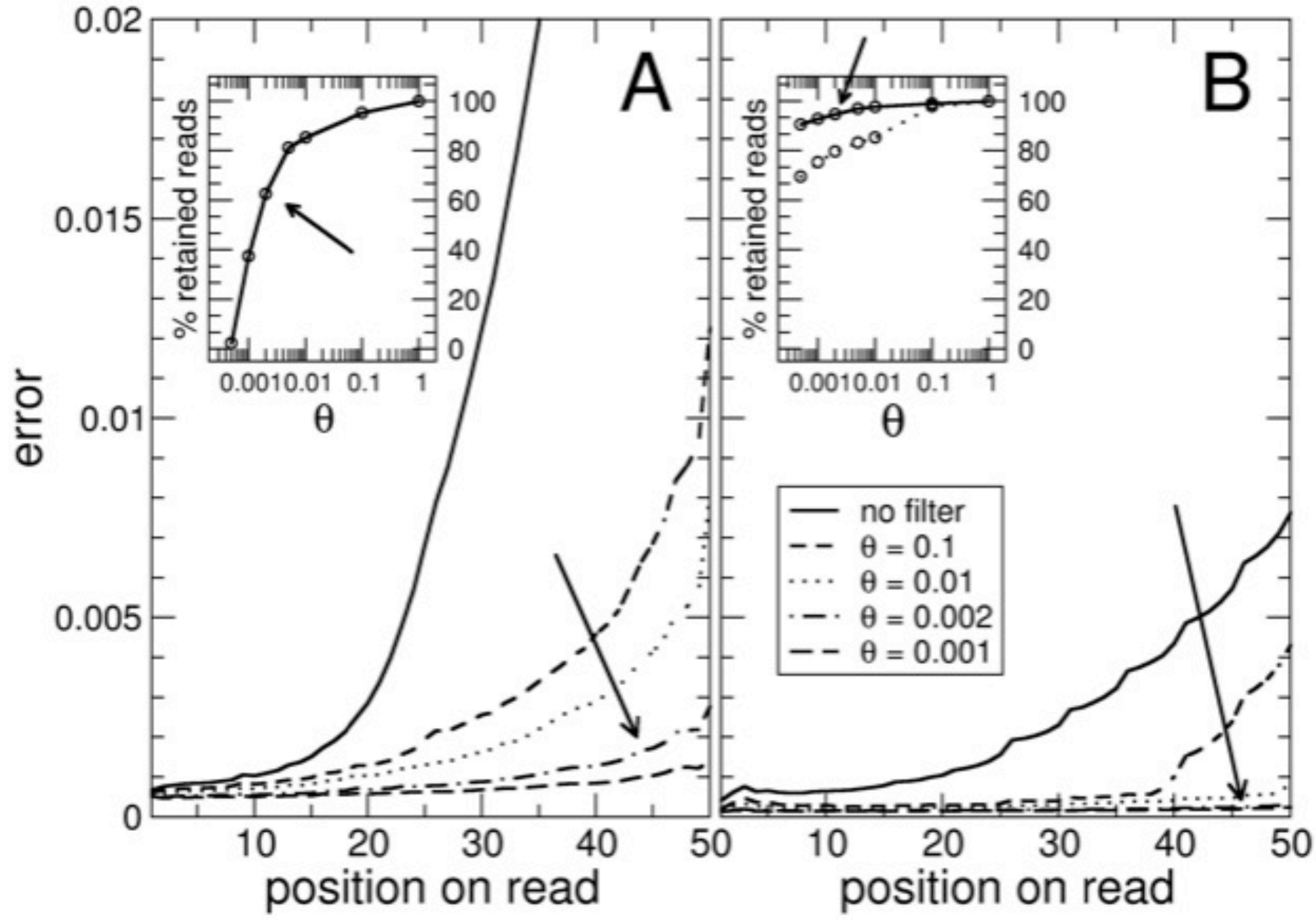

FIG. A1: Average error on reads, computed with base qualities. Panel A: first dataset; panel B: second dataset. The average error increases greatly towards the end of the reads (solid lines). The second dataset was less noisy. Different filtering strategies were tested: only the reads whose average error was below a threshold θ were accepted. More stringent thresholds decrease the errors on the reads (small dashed, dotted, dot-dashed and dashed lines). The insets show the fraction of reads retained after the filtering process (using a threshold θ=0.2%) and retaining 66% of the reads in the first dataset and 95% of the second dataset.

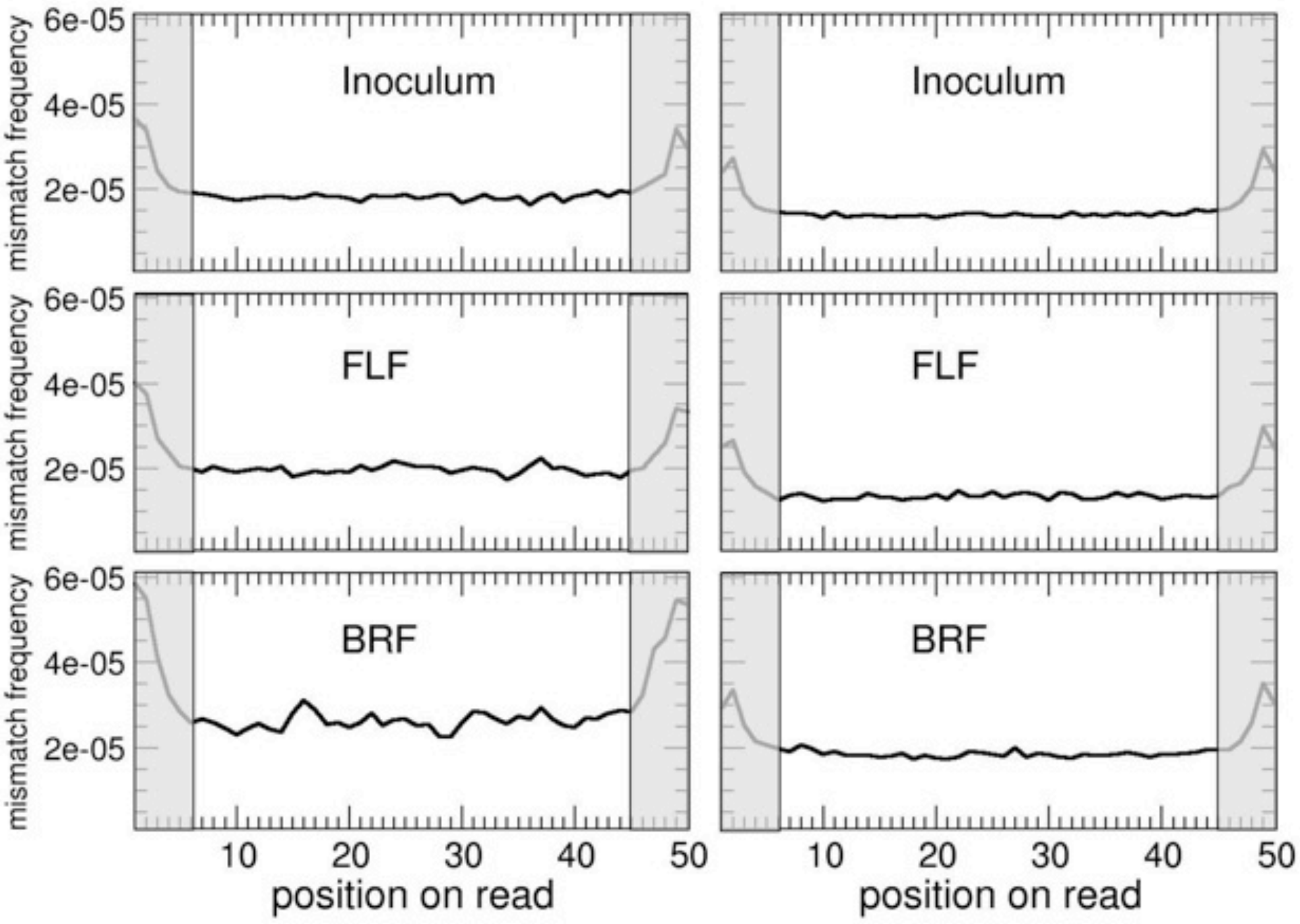

FIG. A2: Distribution of mismatches to the reference genome on the reads after alignment. Left column: first dataset, right column: second dataset. The curves are largely flat, indicating an even distribution of mismatches over the reads, apart for a mild increase towards the edges of the reads, possibly due to reads containing insertions and deletions. We kept only data coming from the flat region of the curve, *i.e.* nucleotides from 5 to 45 in each aligned read.